# A space-fractional cable equation for the propagation of action potentials in myelinated neurons

Corina S. Drapaca[a], Sahin Ozdemir[a], Elizabeth A. Proctor[b,c]

[a]*Department of Engineering Science and Mechanics, Pennsylvania State University, University Park PA 16802, USA*

[b]*Departments of Neurosurgery and Pharmacology, Pennsylvania State College of Medicine, Hershey PA 17033, USA*

[c]*Department of Biomedical Engineering, Pennsylvania State University, University Park PA 16802, USA*

**Abstract**

Myelinated neurons are characterized by the presence of myelin, a multilaminated wrapping around the axons formed by specialized neuroglial cells. Myelin acts as an electrical insulator and therefore, in myelinated neurons, the action potentials do not propagate within the axons but happen only at the nodes of Ranvier which are gaps in the axonal myelination. Recent advancements in brain science have shown that the shapes, timings, and propagation speeds of these so-called saltatory action potentials are controlled by various biochemical interactions among neurons, glial cells, and the extracellular space. Given the complexity of brain's structure and processes, the work hypothesis made in this paper is that non-local effects are involved in the optimal propagation of action potentials. A space-fractional cable equation for the action potentials propagation in myelinated neurons is proposed that involves spatial derivatives of fractional order. The effects of non-locality on the distribution of the membrane potential are investigated using numerical simulations.

## 1- Introduction

The human brain is made of billions of neurons and glial cells separated by the extracellular space (ECS) that facilitates a huge number of cellular interconnections needed for the proper functionality of brain. Neurons are the most important brain cells for the cognitive functions of the brain since they provide brain's functional abilities by transmitting information throughout the body via action potentials. Understanding the fundamental science that governs neuronal processes is crucial for the development of reliable diagnoses and treatments of brain diseases.

While a lot of progress has been made in neuroscience especially in the last few decades when the advancements in technology allowed for much improved observations of cellular and molecular phenomena in the living brain, important knowledge about neuronal processes that could translate into successful treatments of some brain diseases with increasing incidence such as epilepsy, multiple sclerosis and Alzheimer's disease continues to be missing. The main difficulty in achieving this much needed knowledge is the complex nature of neurons and their interactions. Neuronal complexity comes from the fractal-like branched structure of neurons and the highly non-linear processes taking place inside and outside the neurons [1]. Neurons have the following characteristics specific to complex systems: flexibility, adaptability, emergence and noise [1,2]. Of particular interest are emergence and noise. Emergence is a spontaneous event that cannot be predicted from knowledge of the structure and interactions of neurons [1]. According to [3], (membrane) noise is the collection of random fluctuations in the potential of the neuronal membrane which is responsible for action potentials. These fluctuations are caused by the Brownian motion of ions and electrons due to thermal variations (Gaussian white noise), discrete nature of the electric currents through the membrane (shot noise), and conductance changes due to the random opening and closing of ion channels. The noise, emergence and the connections between them represent strong non-linear dynamics which are non-local in space and time [2]. Fluctuation effects at the pico- and nanometric length scales of the ion channels and the ECS cannot be neglected. Dinariev [4,5] proved mathematically that thermal fluctuations can lead to non-local hydrodynamics. Also, at these length scales the long-range interactions among ions and between ions and water molecules are significant [6,7]. Thanks to water's network of intermolecular hydrogen bonds, nanometric dipolar correlations among the water molecules develop at nanoscales which is a non-local dielectric feature of water [8]. Adding ions to water produces solvation shells and experiments and molecular dynamics simulations suggest that the ions and their first hydration shells behave as colloidal suspensions in pure liquid water [9] which can have enhanced long-range interactions at nanoscales.

Another noise-inducing process that could lead to spatio-temporal non-locality is anomalous diffusion of ions described as fractional Brownian motion (fractional Gaussian noise) or Lévy flights (Lévy-stable noise) [10]. For instance, non-locality in neuronal dynamics may be facilitated by the ECS that surrounds the extensively interconnected brain cells (figure 1(a)). The ECS is a collection of narrow channels (20-60 nm width) with convoluted geometries that are filled with interstitial fluid, free and fixed long-chain macromolecules that form the extracellular matrix, and ions involved in cellular resting, action potentials and release of transmitter molecules [11,12]. This space is a constantly fluctuating environment since action potentials and some biochemical interactions among neurons, glial cells and blood capillaries are mediated by the ECS. Diffusion of ions through the ECS might be anomalous due to various factors such as 1) the geometry of the ECS, 2) dead spaces caused by local enlargements of the ECS, voids or glial wrapping around cells, 3) obstruction by the extracellular matrix, 4) biding sites on cell membranes of the extracellular matrix, and 5) the presence of fixed negative charges on the extracellular matrix [12] (figure 1(b)). The fixed negative charges regulate ion mobility [13]. Glial cells, especially astrocytes, also control the ion and water flow in the ECS [14,15]. In particular, the stellate-shaped astrocytes have numerous specialized endfeet linking with other glial cells, neurons, and brain-fluid barriers that are endowed with ion and water channels for the regulation of ion and water homeostasis. For example, the regulation of the extracellular potassium concentration[*] by astrocytes is achieved through two separate processes: 1) an active uptake, accumulation and release of potassium, and 2) passive spatial buffering where the uptake at one location is coupled to the release at a different location. Since both processes cause astrocyte swelling, an intricate process of volume regulation is activated in the astrocyte that will release ion and water in the ECS [14]. We suspect that the tight control of the ion and water movement by astrocytes may also contribute to the ECS non-locality.

An additional source of spatio-temporal non-locality could be the anomalous diffusion of ions through the myelin sheath of neurons. In a myelinated neuron the axon is made of myelinated regions separated by the nodes of Ranvier (figure 2). Myelin is a multilaminated wrapping around the axons made of layers of glial plasma membrane. Oligodendrocytes are specialized glial cells that produce myelin and induce (through processes regulated by neurons and astrocytes) the clustering of voltage-gated sodium channels at the nodes of Ranvier and the aggregation of fast voltage-gated potassium channels in the myelin sheath [16,17]. A few slow potassium channels might also exist in the nodes of Ranvier [18]. Some experiments in cell cultures and in vivo have shown that nodal-like clusters appear just before the deposition of myelin along axons begins [19]. During an action potential, sodium flows inside the neuron at the node of Ranvier, and potassium is released into the periaxonal space which is the extracellular space between the axon and the myelin. To avoid the accumulation of potassium and osmotically driven water in the periaxonal space, potassium diffuses through the myelin's potassium channels and the gap junctions connecting the myelin layers and is removed by the astrocytes endfeet that form gap junctions with the outer most myelin layer [14]. Thus, in myelinated neurons, the action potentials do not propagate within the axons but happen only at the nodes of Ranvier. Experiments performed on the plasma membrane of a cultured human embryonic kidney cell showed that the voltage-gated potassium channel Kv2.1 displays anomalous diffusion [20]. Given that the Kv2.1 channel shows similar clustering patterns in the membranes of cultured human embryonic kidney cells and native neurons [20], it is possible that anomalous diffusion of potassium may also happen in the myelin sheath of neurons.

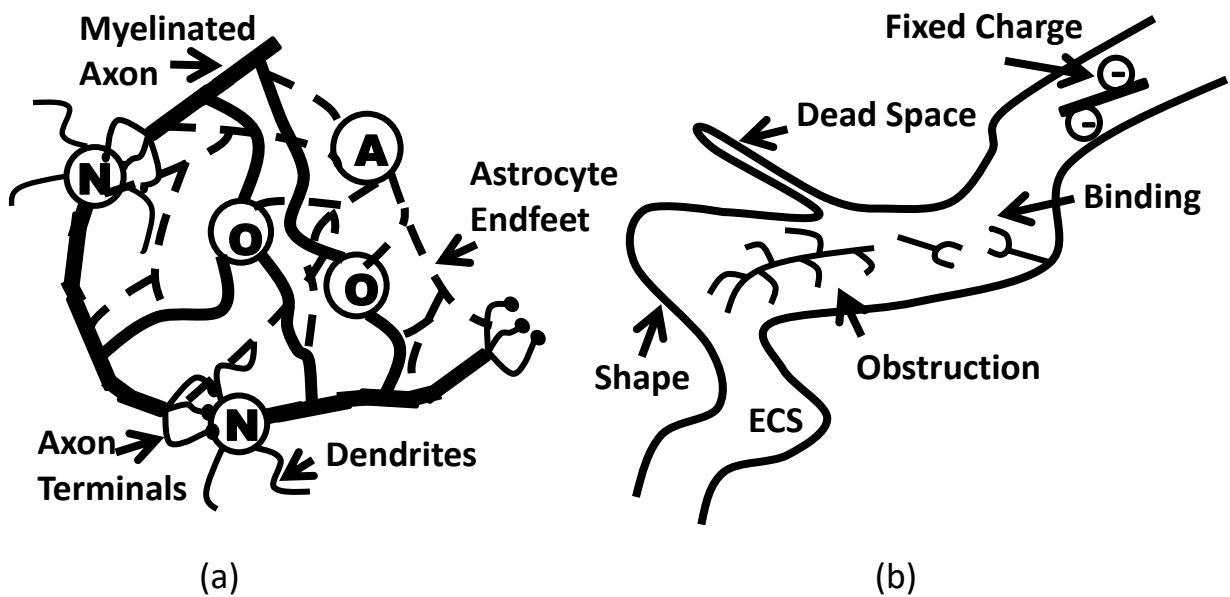

**Figure 1.** **(a) Schematic representation of the interactions among myelinated neurons (N), astrocytes (A) and myelinating oligodendrocytes (O). Astrocytes have numerous specialized connections (endfeet) to other astrocytes, the soma, dendrites and axons of neurons, the soma and processes of oligodendrocytes, and the brain fluid barriers. Neurons and astrocytes control the production and assembly of the myelin sheath around the axons by the oligodendrocytes. One oligodendrocyte can myelinate many axonal segments belonging to different neurons. Interactions among astrocytes and oligodendrocytes are controlled by gap junctions, while neuronal interactions happen through electrical and chemical synapses. (adapted from [16]) (b) Schematic representation of the diffusion barriers of the ECS: geometry, dead space microdomains, obstruction by the extracellular matrix, biding sites, and fixed negative charges on the extracellular matrix (adapted from [12]).**

---

[*] A disruption in the potassium regulation is a sign of neuronal network dysfunction [20]. For instance, during an epileptic seizure, the extracellular potassium concentration increases.

In this paper we assume that the spatio-temporal non-locality due to, among other factors, long-range interactions of ions and water molecules at nanoscales and the anomalous diffusion of ions through the myelin sheath and the ECS, has a non-negligible effect on the propagation of action potentials in myelinated neurons. We propose a mathematical model of the action potential propagation that accounts for spatial non-local effects. To model spatial non-locality, we use fractional calculus which has been successfully employed to describe spatio-temporal non-locality in various complex systems (see for instance [21,22] and references within, and [23] for anomalous diffusion). We generalize the classic cable equation that models the spatio-temporal propagation of action potentials in neurons. We derive the model for mature myelinated neurons which have backpropagating action potentials [24]. In the numerical simulations we will consider the unidirectional propagation of the action potentials as shown in [47] as well as the bidirectional case. We start by introducing the non-local voltage as a convolution whose kernel is a decaying power law of fractional order. This kernel function can model the Lévy flights/fractional Brownian motion characterizing anomalous diffusion [10,23] and the superposition of exponentially decaying screened Coulomb potentials representing long-range electrostatic interactions of ions [25]. An optimal approximation of a decaying power law by exponentials can be found in [26]. We further derive a non-local cable equation using fractional order derivatives and the same steps as in [27] where the classic cable equation is presented. By replacing the spatial derivatives of integer order in the cable equation (which is a one-dimensional parabolic partial differential equation) with spatial fractional Caputo derivatives of fractional order we obtain a one-dimensional integro- differential equation. We notice that such an approach has not been proposed in the literature until now. The fractional cable equation proposed in [28,29] uses a temporal fractional order derivative to introduce a diffusion operator in the classic cable equation. This equation models only temporal anomalous electro-diffusion in neurons, not spatial non-locality.

The structure of the paper is as follows. Some mathematical preliminaries needed in the paper are given in section 2. The proposed mathematical model is presented in section 3, together with a numerical solution to the non-local cable equation coupled with the Hodgkin-Huxley equations. Numerical simulations are shown in section 4. Lastly, the paper ends with a section of conclusions and further work.

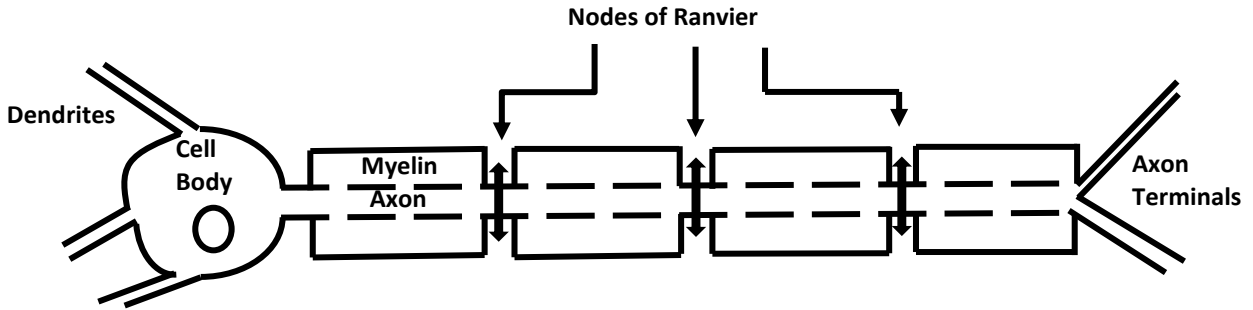

**Figure 2.** Schematic of a myelinated neuron: the information travels from the cell body to the terminal through the axon made of myelinated regions separated by the nodes of Ranvier. The information is transmitted through action potentials which happen only at the nodes of Ranvier and are represented by vertical double arrows. The action potentials propagate between the cell body and the axon terminals (bidirectional propagation).

## 2- Preliminaries

In this section we present some fractional calculus results that will be used in the paper. The mathematical concepts we need in this paper are taken from [30-34].

**Definition 2.1:** 1. Let $f \in L^1(a,b)$ and $\alpha > 0$. The *left-sided Riemann-Liouville fractional integral of order $\alpha$* is:

$$I_{a+}^{\alpha} f(x) = \frac{1}{\Gamma(\alpha)} \int_a^x \frac{f(y)}{(x-y)^{1-\alpha}} dy \tag{1}$$

and the *right-sided Riemann-Liouville fractional integral of order $\alpha$* is:

$$I_{b-}^{\alpha}f(x) = \frac{1}{\Gamma(\alpha)} \int_x^b \frac{f(y)}{(y-x)^{1-\alpha}} dy \qquad (2)$$

where $\Gamma(s) = \int_0^\infty t^{s-1}e^{-t}dt$ is the gamma function. In particular, $I_{a+}^0 f(x) = I_{b-}^0 f(x) = f(x)$.

2. If $f:[a,b] \to \mathbf{R}$ is $n$-times differentiable on $(a,b) \subset \mathbf{R}$ with $f^{(n)} \in L^1(a,b)$ and $n-1 < \alpha < n$, $n = 1,2,3 ...$, then the *left-sided Caputo fractional derivative of order $\alpha$* is:

$$D_{a+}^{\alpha}f(x) = \frac{1}{\Gamma(n-\alpha)} \int_a^x \frac{f^{(n)}(y)}{(x-y)^{\alpha+1-n}} dy = I_{a+}^{n-\alpha} f^{(n)}(x) \qquad (3)$$

and the *right-sided Caputo fractional derivative of order $\alpha$* is:

$$D_{b-}^{\alpha}f(x) = \frac{(-1)^n}{\Gamma(n-\alpha)} \int_x^b \frac{f^{(n)}(y)}{(y-x)^{\alpha-n+1}} dy = (-1)^n I_{b-}^{n-\alpha} f^{(n)}(x) \qquad (4)$$

where $f^{(n)}$ is the $n^{\text{th}}$-order derivative of $f$. In particular, $D_{a+}^n f(x) = f^{(n)}(x)$, and $D_{b-}^n f(x) = (-1)^n f^{(n)}(x)$.

3. The left (right)-sided Caputo fractional derivative of order $n\alpha$, with $n-1 < n\alpha < n, n = 2,3, ...$, is said to be a left(right)-sided *sequential Caputo fractional derivative* if:

$$D_{a+}^{n\alpha} f(x) = D_{a+}^{\alpha}\left(D_{a+}^{(n-1)\alpha}\right) f(x), \quad D_{b-}^{n\alpha} f(x) = D_{b-}^{\alpha}\left(D_{b-}^{(n-1)\alpha}\right) f(x) \qquad (5)$$

For $0 < \alpha < 1$, $d(D_{a+}^{\alpha}f(x))/dx = D_{a+}^{\alpha+1}f(x)$ if $f(a) = 0$, and $d(D_{b-}^{\alpha}f(x))/dx = -D_{b-}^{\alpha+1}f(x)$ if $f(b) = 0$.

**Proposition 2.1:** 1. If $\alpha > 0, Re(\beta) > 0$ then:

$$I_{a+}^{\alpha}(x-a)^{\beta-1} = \frac{\Gamma(\beta)}{\Gamma(\alpha+\beta)}(x-a)^{\alpha+\beta-1}, x > a$$

$$I_{b-}^{\alpha}(b-x)^{\beta-1} = \frac{\Gamma(\beta)}{\Gamma(\alpha+\beta)}(b-x)^{\alpha+\beta-1}, x < b \qquad (6)$$

$$D_{a+}^{\alpha}(x-a)^{\beta-1} = \frac{\Gamma(\beta)}{\Gamma(\beta-\alpha)}(x-a)^{\beta-\alpha-1}, x > a, 0 < \alpha < 1$$

$$D_{b-}^{\alpha}(b-x)^{\beta-1} = \frac{\Gamma(\beta)}{\Gamma(\beta-\alpha)}(b-x)^{\beta-\alpha-1}, x < b, 0 < \alpha < 1 \qquad (7)$$

$$D_{a+}^{\alpha}c = D_{b-}^{\alpha}c = 0, \text{ c constant} \qquad (8)$$

2. If $n-1 < \alpha \le n$ for $n = 1, 2, 3 ...$ and the function $f$ has the property that there exist $p > \mu \ge -1$ and a continuous function $g$ such that $f^{(n)}(x) = x^p g(x)$ for $x > a$, then:

$$D_{a+}^{\alpha} I_{a+}^{\alpha} f(x) = f(x),$$

$$I_{a+}^{\alpha} D_{a+}^{\alpha} f(x) = f(x) - \sum_{j=0}^{n-1} f^{(j)}(a+) \frac{x^j}{j!} \qquad (9)$$

3. If $0 < \alpha \le 1$ and the function $f$ has continuous sequential Caputo derivatives $D^{k\alpha}f$ for $k = 0, 1, ..., n+1$, then the following generalized Taylor's formulas hold for $a < x < b$:

$$f(x) = \sum_{k=0}^{n} \frac{D_{a+}^{\alpha} f(a)}{\Gamma(k\alpha + 1)} (x - a)^{k\alpha} + \frac{D_{0+}^{(n+1)\alpha} f(\xi)}{\Gamma((n+1)\alpha + 1)} (x - a)^{(n+1)\alpha}, a \leq \xi \leq x$$

$$f(x) = \sum_{k=0}^{n} \frac{D_{b-}^{\alpha} f(b)}{\Gamma(k\alpha + 1)} (b - x)^{k\alpha} + \frac{D_{b-}^{(n+1)\alpha} f(\xi)}{\Gamma((n+1)\alpha + 1)} (b - x)^{(n+1)\alpha}, x \leq \xi \leq b$$

(10)

The following representations are also valid:

$$f(x) = f(x_0) + \frac{D_{a+}^{\alpha} f(x_0)}{\Gamma(\alpha + 1)} ((x - a)^{\alpha} - (x_0 - a)^{\alpha}) + R_{2\alpha}$$

$$f(x) = f(x_0) + \frac{D_{b-}^{\alpha} f(x_0)}{\Gamma(\alpha + 1)} ((b - x)^{\alpha} - (b - x_0)^{\alpha}) + \tilde{R}_{2\alpha}$$

(11)

where $R_{2\alpha}$, $\tilde{R}_{2\alpha}$ are reminder terms.

## 3- Mathematical Model

In this section we introduce a generalization of the cable equation using spatial derivatives of fractional order. The bidirectional propagation of action potentials between the cell body and the axon terminals is modeled using left- and right-sided fractional order operators.

**Definition 3.1:** The *non-local voltage of order* $\boldsymbol{\alpha}(t) = (\alpha_1(t), \alpha_2(t), \alpha_3(t))$, $\alpha_i(t) > 0, i = 1,2,3$ between two points in $\boldsymbol{R}^3$ of position vectors $\boldsymbol{0} = (0, 0, 0)$ and $\boldsymbol{x} = (x_1, x_2, x_3) \in (\boldsymbol{0}, \boldsymbol{R}), \boldsymbol{R} = (R_1, R_2, R_3)$ is:

$$\begin{aligned}
V(\boldsymbol{x}, t) = &-\frac{p}{L_1^{\alpha_1(t)-1} \Gamma(\alpha_1(t))} \int_0^{x_1} (x_1 - y_1)^{\alpha_1(t)-1} E_1(\boldsymbol{y}, t) dy_1 \\
&-\frac{p}{L_2^{\alpha_2(t)-1} \Gamma(\alpha_2(t))} \int_0^{x_2} (x_2 - y_2)^{\alpha_2(t)-1} E_2(\boldsymbol{y}, t) dy_2 \\
&-\frac{p}{L_3^{\alpha_3(t)-1} \Gamma(\alpha_3(t))} \int_0^{x_3} (x_3 - y_3)^{\alpha_3(t)-1} E_3(\boldsymbol{y}, t) dy_3 \\
&-\frac{q}{L_1^{\alpha_1(t)-1} \Gamma(\alpha_1(t))} \int_{x_1}^{R_1} (y_1 - x_1)^{\alpha_1(t)-1} E_1(\boldsymbol{y}, t) dy_1 \\
&-\frac{q}{L_2^{\alpha_2(t)-1} \Gamma(\alpha_2(t))} \int_{x_2}^{R_2} (y_2 - x_2)^{\alpha_2(t)-1} E_2(\boldsymbol{y}, t) dy_2 \\
&-\frac{q}{L_3^{\alpha_3(t)-1} \Gamma(\alpha_3(t))} \int_{x_3}^{R_3} (y_3 - x_3)^{\alpha_3(t)-1} E_3(\boldsymbol{y}, t) dy_3
\end{aligned}$$

(12)

where $\boldsymbol{E}(\boldsymbol{y}, t) = (E_1(\boldsymbol{y}, t), E_2(\boldsymbol{y}, t), E_3(\boldsymbol{y}, t))$ is the electric field vector, $t \geq 0$ is a non-dimensional time, and $L_i, i = 1, 2, 3$ are characteristic lengths. The positive constants $p$ and $q$ satisfy the constraint $p + q = 1$. As in the classic approach, we take $V(\boldsymbol{0}) = 0$ for mathematical convenience.

For now, we do not consider time-dependent spatial non-local effects. In addition, since the axonal branches are long and narrow, the action potentials depend only on one spatial variable [27] and thus the problem is one-dimensional.

Therefore, we model the axon as a circular cylinder of constant radius $r$ and one characteristic length $L$, the length of the internodal region, and assume that:

$$\alpha_1(t) = \alpha_2(t) = \alpha_3(t) = \alpha = constant, 0 < \alpha \leq 1 \tag{13}$$

In this case, the element of path along the integration of $\vec{E}$ (a generalization of a measure proposed in [35]):

$$d\boldsymbol{y}_{\boldsymbol{\alpha}(t)} = \left(\frac{1}{\Gamma(\alpha_1(t))} y_1^{\alpha_1(t)-1} dy_1, \frac{1}{\Gamma(\alpha_2(t))} y_2^{\alpha_2(t)-1} dy_2, \frac{1}{\Gamma(\alpha_3(t))} y_3^{\alpha_3(t)-1} dy_3\right)$$

reduces to:

$$d\boldsymbol{y}_{\boldsymbol{\alpha}} = \frac{y^{\alpha-1}}{\Gamma(\alpha)} d\boldsymbol{y}.$$

We assume further that the electric field is uniform along the axon and the standard convention of current direction is valid: the membrane and synaptic current are positive when they are outward, and the electrode currents are positive when they are inward. Then formula (12) becomes:

$$\begin{aligned} V(x,t) &= -\frac{Ep}{L^{\alpha-1}} \int_0^x d(x-y)_\alpha - \frac{Eq}{L^{\alpha-1}} \int_x^R d(y-x)_\alpha = -\frac{Ep}{L^{\alpha-1}} \int_0^x d\left[\frac{(x-y)^\alpha}{\Gamma(\alpha+1)}\right] \\ &\quad - \frac{Eq}{L^{\alpha-1}} \int_x^R d\left[\frac{(y-x)^\alpha}{\Gamma(\alpha+1)}\right] = \frac{E}{L^{\alpha-1}} \frac{px^\alpha - q(R-x)^\alpha}{\Gamma(\alpha+1)} \end{aligned} \tag{14}$$

By comparing formulas (14) and (6) we notice that:

$$V(x,t) = pI_{0+}^\alpha \left(\frac{E}{L^{\alpha-1}}\right) - qI_{R-}^\alpha \left(\frac{E}{L^{\alpha-1}}\right)$$

From formula (14) we further get (for an equally-spaced discretization of $[0, R]$ with step size $\Delta x$ and the number of nodes chosen such that, when $\alpha = 1$, the following formula reduces to the classic forward approximation of the first order derivative):

$$\Delta V(x,t) = V(x+\Delta x, t) - V(x,t) = \frac{E}{L^{\alpha-1}\Gamma(\alpha+1)}((x+\Delta x)^\alpha - x^\alpha) \tag{15}$$

On the other hand, using formulas (11) and the fact that $p + q = 1$ give:

$$\begin{aligned} \Delta V(x,t) &= (p+q)V(x+\Delta x, t) - (p+q)V(x,t) \\ &\approx \frac{pD_{0+}^\alpha V(x,t) + q(-1)^\alpha D_{R-}^\alpha V(x,t)}{\Gamma(\alpha+1)}((x+\Delta x)^\alpha - x^\alpha) \end{aligned} \tag{16}$$

where the Caputo fractional derivatives are taken with respect to the spatial variable $x$.

Thus, as $\Delta x \to 0$, formulas (15) and (16) yield:

$$pD_{0+}^\alpha V(x,t) + q(-1)^\alpha D_{R-}^\alpha V(x,t) = \frac{E}{L^{\alpha-1}} \tag{17}$$

The uniformity of the current density in every cross-sectional area of the cylindrical neuron gives [27]:

$$E = \frac{r_L I}{\pi r^2} \tag{18}$$

where $r_L$ is the intracellular resistance and $I$ is the electric current. By combining formulas (17) and (18) we get the expression of the longitudinal current:

$$I_L = -\frac{\pi r^2 L^{\alpha-1}}{r_L} [p\, D_{0+}^\alpha V(x,t) + q(-1)^\alpha D_{R-}^\alpha V(x,t)] \tag{19}$$

where the negative sign comes from the current sign convention. Also, replacing formula (18) into formula (15) yields the Ohm's law:

$$\Delta V(x,t) = R_\alpha I$$

for the generalized electric resistance:

$$R_\alpha = \frac{r_L}{\pi r^2 L^{\alpha-1} \Gamma(\alpha+1)} ((x+\Delta x)^\alpha - x^\alpha) \tag{20}$$

Further we denote by:

$$\tilde{r}_L = \frac{r_L}{L^{\alpha-1}}$$

As in the classic case [1], we introduce the following currents:

$$\begin{aligned} I_m &= 2\pi r\, \Delta x\, i_m \\ I_e &= 2\pi r\, \Delta x\, i_e \\ I_c &= 2\pi r\, \Delta x\, c_m \frac{\partial V}{\partial t} \end{aligned} \tag{21}$$

where $I_m$ and $I_e$ are the membrane and external currents, respectively, and $i_m$ and $i_e$ are their corresponding currents per unit area. The capacitor current is $I_c$ and the specific membrane capacitance is $c_m$.

If we replace now formulas (19) and (21) in the Kirchhoff's law in the element shown in figure 3b:

$$I_c + I_m - I_{L|left} + I_{L|right} - I_e = 0$$

we obtain:

$$c_m \frac{\partial V}{\partial t} = i_e - i_m$$
$$+ \frac{r}{2\tilde{r}_L} \frac{[p\, D_{0+}^\alpha V(x,t) + q(-1)^\alpha D_{R-}^\alpha V(x,t)]_{|right} - [p\, D_{0+}^\alpha V(x,t) + q(-1)^\alpha D_{R-}^\alpha V(x,t)]_{|left}}{\Delta x} \tag{22}$$

As $\Delta x \to 0$:

$$\frac{[p\, D_{0+}^\alpha V(x,t) + q(-1)^\alpha D_{R-}^\alpha V(x,t)]_{|right} - [p\, D_{0+}^\alpha V(x,t) + q(-1)^\alpha D_{R-}^\alpha V(x,t)]_{|left}}{\Delta x}$$
$$\approx \frac{\partial}{\partial x} [p\, D_{0+}^\alpha V(x,t) + q(-1)^\alpha D_{R-}^\alpha V(x,t)]$$

which replaced in equation (22) and letting $\Delta x \to 0$ give the expression of the *non-local, space-fractional cable equation*:

$$c_m \frac{\partial V}{\partial t} = \frac{r}{2\tilde{r}_L} \left[ p \frac{\partial}{\partial x} (D_{0+}^\alpha V) + q(-1)^\alpha \frac{\partial}{\partial x} (D_{R-}^\alpha V) \right] + i_e - i_m \tag{23}$$

If $V(0,t) = V(R,t) = 0$, equation (23) can be written in the equivalent form:

$$c_m \frac{\partial V}{\partial t} = \frac{r}{2\tilde{r}_L}[pD_{0+}^{\alpha+1}V + q(-1)^{\alpha+1}D_{R-}^{\alpha+1}V] + i_e - i_m$$

Equation (23) is a space-fractional reaction-diffusion equation. For $p = q = 1/2$ and $\alpha = 1$, equation (23) reduces to the classic cable equation [27]:

$$c_m \frac{\partial V}{\partial t} = \frac{r}{2r_L} \frac{\partial^2 V}{\partial x^2} + i_e - i_m \qquad (24)$$

A numerical discretization of the second order spatial derivative in equation (24) shows that in myelinated neurons the voltage at node $n$ depends only on the voltages at the adjacent nodes of Ranvier $n-1$ and $n+1$. However, a numerical discretization of the spatial fractional derivatives in equation (23) based on the Grünwald-Letnikov formula (see later) shows that the voltage at node $n$ depends on the voltages at all the previous nodes $0, 1, \dots, n-1$, and all the following nodes $n+1, n+2, \dots n+N$, where $N$ is the total number of nodes of an axon.

Near the resting potential $V_{rest}$ the membrane current per unit area can be approximated as [27]:

$$i_m = \frac{V - V_{rest}}{r_m} := \frac{v}{r_m}$$

where $v = V - V_{rest}$ is the voltage relative to the resting potential, and $r_m$ is the specific membrane resistance. Since $V_{rest}$ is constant it follows from formula (8) that:

$$D_{0+}^{\alpha}V = D_{0+}^{\alpha}v, \qquad D_{R-}^{\alpha}V = D_{R-}^{\alpha}v$$

Also:

$$\frac{\partial V}{\partial t} = \frac{\partial v}{\partial t}$$

Thus, equation (23) becomes:

$$\tau_m \frac{\partial v}{\partial t} = \lambda^{\alpha+1}\left[p \frac{\partial}{\partial x}(D_{0+}^{\alpha}v) + q(-1)^{\alpha}\frac{\partial}{\partial x}(D_{R-}^{\alpha}v)\right] + r_m i_e - v \qquad (25)$$

where we denoted by:

$$\tau_m = r_m c_m, \lambda^{\alpha+1} = \frac{rr_m}{2\;r_L}L^{\alpha-1}$$

Here $\lambda$ is a generalized electrotonic length that shows the role of parameter $\alpha \in (0,1)$ as a non-local corrector of the relationship among the geometric parameters $r, L$ and the resistances $r_m, r_L$ of a neuron. We notice that equation (23) remains valid for a time-varying $\alpha(t)$, thus allowing the modeling of dynamic effects on $r, L, r_m$, and $r_L$ through varying $\alpha$. This might provide relevant insights into how aging and neuro-glial processes control the myelin assembly so that the timing, speed and shape of the action potentials are optimized. For instance, by relating $\alpha(t)$ to aging we might be able to understand the variations in the lengths and diameters of the internodal regions and nodes along one axon or between two neurons [36-38]. It may also be possible to link parameter $\alpha(t)$ to a mechanical model of swelling to account for the change in the axon's diameter during an action potential [39], or to the diffusion of the extracellular potassium to account for changes in the membrane resistance [18].

In the following sub-section, we will present a numerical solution to equation (25) coupled with the Hodgkin-Huxley equations. The problem is stated and solved in the internodal region (figure 3a).

### *3-1- Dynamic Problem*

We consider first the case of unidirectional propagation of action potentials from the cell body to the axon terminals and thus take $p = 1, q = 0$ in equation (25). We assume a leaky internodal region with one isopotential node described by the modified Hodgkin-Huxley model proposed in [40]. The reason why we chose the Hodgkin-Huxley model instead of the more commonly used (see for instance [41,42]) model by Frankenhaeuser and Huxley [43] is that the Frankenhaeuser-Huxley model provides an incorrect description of the sodium currents [3]. The membrane potential of the internodal region is solution to the following initial boundary value problem [47]:

$$\tau_m \frac{\partial v}{\partial t} = \lambda^{\alpha+1} \frac{\partial}{\partial x}(D_{0+}^\alpha v) - v, \quad x \in (0, L) \tag{26}$$

$$v(x,0) = 0, v(0,t) = 0, v(L,t) = f(t)$$

where $f(t) = V(t) - V_{rest}$ and $V(t)$, the membrane potential of the node, is solution of the modified Hodgkin-Huxley model [40] reproduced here for completeness.

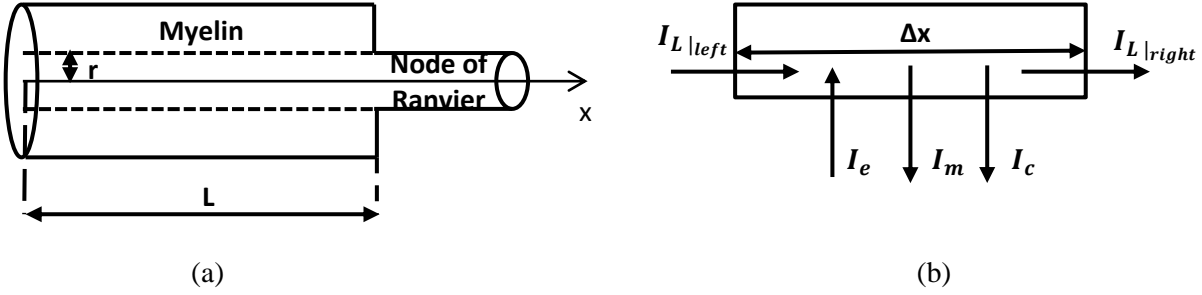

(a) (b)

**Figure 3.** (a) Schematic of a part of a neuron made of an internodal (myelinated) region of length $L$ and a node of Ranvier. The radius of the neuron is denoted by $r$. (b) The electric currents entering (the longitudinal current $I_{L|left}$ and external current $I_e$) and exiting ( the longitudinal current $I_{L|right}$, the membrane current $I_m$, and the capacitor current $I_c$) a neuronal segment of length $\Delta x$.

$$\begin{aligned}
C_m \frac{dV}{dt} &= I - (G_{Na}m^3h + G_{NaL})(V - E_{Na}) - (G_K n^4 + G_{KL})(V - E_K) \\
&\quad - G_{ClL}(V - E_{Cl}) \\
\frac{dm}{dt} &= \alpha_m(1-m) - \beta_m m \\
\frac{dn}{dt} &= \alpha_n(1-n) - \beta_n n \\
\frac{dh}{dt} &= \alpha_h(1-h) - \beta_h h \\
V(0) &= V_{rest} \\
m(0) &= \frac{\alpha_m(V(0))}{\alpha_m(V(0)) + \beta_m(V(0))} \\
n(0) &= \frac{\alpha_n(V(0))}{\alpha_n(V(0)) + \beta_n(V(0))} \\
h(0) &= \frac{\alpha_h(V(0))}{\alpha_h(V(0)) + \beta_h(V(0))} \\
\alpha_m &= \frac{0.32(V + 54)}{1 - \exp(-0.25(V + 54))}, \beta_m = \frac{0.28(V + 27)}{\exp(0.2(V + 27)) - 1} \\
\alpha_n &= \frac{0.032(V + 52)}{1 - \exp(-0.2(V + 52))}, \beta_n = 0.5 \exp\left(-\frac{V + 57}{40}\right) \\
\alpha_h &= 0.128 \exp\left(-\frac{V + 50}{18}\right), \beta_h = \frac{4}{1 + \exp(-0.2(V + 27))}
\end{aligned} \tag{27}$$

Above $I$ is an externally applied current per unit area, $E_{Na}, E_K$, and $E_{Cl}$ are the reverse potentials, $G_{Na}, G_K, G_{NaL}, G_{KL}$, and $G_{ClL}$ are respectively the voltage-gated maximal conductances of $Na^+$ and $K^+$, and the leak conductances of $Na^+, K^+$, and $Cl^-$. Lastly, the gating variables $m, n$, and $h$ represent the activations of the $Na^+$ and $K^+$ channels and the inactivation of the $Na^+$ channel, respectively.

Let:
$$w(x,t) = v(x,t) - \tilde{v}(x,t) \tag{28}$$

where $\tilde{v}(x,t)$ is solution to the following boundary value problem:
$$\begin{aligned}&\frac{\partial}{\partial x}(D_{0+}^\alpha \tilde{v}) = 0 \\ &\tilde{v}(0,t) = 0 = v(0,t) \\ &\tilde{v}(L,t) = f(t) = v(L,t)\end{aligned} \tag{29}$$

Then the following identities hold:
$$\frac{\partial}{\partial x}(D_{0+}^\alpha w) = \frac{\partial}{\partial x}(D_{0+}^\alpha v), w(0,t) = w(L,t) = 0 \tag{30}$$

The solution to problem (29) can be found as follows. Integrating one the differential equation gives:
$$D_{0+}^\alpha \tilde{v} = c = \text{constant}$$

The above identity can be transformed further into:
$$\tilde{v}(x,t) - \tilde{v}(0+,t) = I_{0+}^\alpha (D_{0+}^\alpha \tilde{v}) = I_{0+}^\alpha c = \frac{cx^\alpha}{\Gamma(\alpha+1)} \tag{31}$$

by using formulas (9) and (6). Finally, using the boundary conditions of problem (29) in formula (31) we find the solution to problem (29):
$$\tilde{v}(x,t) = f(t)\left(\frac{x}{L}\right)^\alpha \tag{32}$$

From the definition of $f(t)$ it follows that $\tilde{v}(x,0) = 0$ which implies that $w(x,0) = 0$. Thus an initial boundary value problem equivalent to problem (26) is:
$$\begin{aligned}&\tau_m \frac{\partial w}{\partial t} = \lambda^{\alpha+1}\frac{\partial}{\partial x}(D_{0+}^\alpha w) - w - \left(f(t) + \tau_m \frac{df(t)}{dt}\right)\left(\frac{x}{L}\right)^\alpha \\ &w(0,t) = w(L,t) = 0 \\ &w(x,0) = 0\end{aligned} \tag{33}$$

We use now the numerical scheme proposed in [44] to discretize the spatial derivatives. The numerical scheme is based on a shifted Grünwald-Letnikov formula that is stable. Let $0 = x_0 < x_1 < \cdots < x_{N-1} < x_N = L$ be an equally spaced discretization of the interval $[0, L]$ of constant step size $\Delta x$. Then the spatially discretized equation (33) with zero Dirichlet boundary conditions is:
$$\begin{aligned}&\frac{\partial w(x_j,t)}{\partial t} = \frac{\lambda^{\alpha+1}}{\tau_m \Delta x^{\alpha+1}}\sum_{i=0}^N b_{ij} w(x_i,t) - \frac{1}{\tau_m} w(x_j,t) - \frac{1}{\tau_m}\left(f(t) + \tau_m \frac{df(t)}{dt}\right)\left(\frac{x_j}{L}\right)^\alpha, \\ &0 \leq j \leq N\end{aligned} \tag{34}$$

with

$$b_{ij} = \begin{cases} \dfrac{(-1)^{j-i+1}\Gamma(\alpha+2)}{\Gamma(j-i+2)\Gamma(\alpha+1-j+i)}, & 0 < j < N, i \leq j+1 \\ 0, & \text{otherwise} \end{cases} \tag{35}$$

System (34) with the zero initial condition of problem (33) together with problem (27) are further solved numerically using Matlab's built-in function **ode15s**. This function solves systems of stiff first order ordinary differential equations by combining a modified linear multistep backward difference formula of order up to 5 and an adaptive step size that changes according to an algorithm that calculates relative and absolute error tolerances [45]. In our simulations we kept the default values of **ode15s** for the relative error tolerance ($10^{-3}$) and the absolute error tolerance ($10^{-6}$). Lastly, the spatial step size $\Delta x$ was chosen such that the following stability condition for system (34) is satisfied:

$$(\alpha + 1)\frac{\lambda^{\alpha+1}}{\tau_m \Delta x^{\alpha+1}}\Delta t \leq 1$$

where $\Delta t$ is the time step.

Now we consider the case of bidirectional propagation of action potentials and solve numerically the following initial boundary value problem for the membrane potential of the internodal region:

$$\tau_m \frac{\partial v}{\partial t} = \lambda^{\alpha+1}\left[p\frac{\partial}{\partial x}(D_{0+}^{\alpha}v) + q(-1)^{\alpha}\frac{\partial}{\partial x}(D_{L-}^{\alpha}v)\right] - v, \quad x \in (0, L) \tag{36}$$

$$v(x,0) = 0, v(0,t) = 0, v(L,t) = f(t)$$

where we took $R = L$ and $f(t)$ was defined earlier. A forward Euler scheme is used to approximate the time derivative, and the fractional derivatives are approximated using the right and left shifted Grünwald-Letnikov formulas [48]:

$$\frac{\partial}{\partial x}\left(D_{0+}^{\alpha}v(x_j t)\right) = \frac{1}{\Delta x^{\alpha+1}}\sum_{i=0}^{j+1} g_i^{\alpha+1} v(x_{j-i+1}, t), 0 < j < N$$

$$\frac{\partial}{\partial x}\left(D_{L-}^{\alpha}v(x_j t)\right) = \frac{-1}{\Delta x^{\alpha+1}}\sum_{i=0}^{N-j+1} g_i^{\alpha+1} v(x_{j+i-1}, t) + \frac{f(t)}{\Gamma(-\alpha)(L - x_j)^{\alpha+1}}, 0 < j < N \tag{37}$$

where we used the boundary conditions from (36) and denoted by:

$$g_i^{\alpha+1} = \frac{(-1)^i \Gamma(\alpha+2)}{\Gamma(i+1)\Gamma(\alpha-i+2)}$$

The stability condition of the numerical scheme for problem (36) is:

$$(\alpha + 1)\frac{\lambda^{\alpha+1}}{\tau_m \Delta x^{\alpha+1}}(p + q|(-1)^{\alpha}|)\Delta t \leq 1$$

which is the same as the stability condition of the numerical scheme used for problem (34) since $p + q = 1$. Problem (27) was solved first using Matlab's built-in function **ode15s**, and its solution was then replaced in scheme (37) which was further implemented in Matlab. The real part of $(-1)^{\alpha}$ was used in the numerical simulations.

## 4- Results

The results for the unidirectional case are published in [47]. The parameters used in our numerical simulations are given in Table 1. For our simulations we chose $L = 1000~\mu m$ which is within the range of values for the internodal region found in the literature (for instance, [46] reports $L = 200 \div 2000~\mu m$ where the lower (higher) values correspond to initial myelination (full maturity) stage [36]). Also, according to the results in [37], a maximum propagation speed of an action potential is reached around a critical value of $1000~\mu m$ for the internodal length. In all the numerical simulations we used a time step $\Delta t = 0.01$. Figures 4-10 present the results for the unidirectional case, while figures 11-14 present the results for the bidirectional case.

**Table 1.** List of parameters with corresponding values and physical units.

| Parameters | Values and Units | Reference |
|---|---|---|
| $V_{rest}$ | -65 mV | [40] |
| $E_{Na}$ | 60 mV | [40] |
| $E_K$ | -88 mV | [40] |
| $E_{Cl}$ | -61 mV | [40] |
| $G_{Na}$ | 0.3 mS/mm$^2$ | [40] |
| $G_K$ | 0.25 mS/mm$^2$ | [40] |
| $G_{NaL}$ | 0.000247 mS/mm$^2$ | [40] |
| $G_{KL}$ | 0.0005 mS/mm$^2$ | [40] |
| $G_{ClL}$ | 0.001 mS/mm$^2$ | [40] |
| $c_m$ | 0.01 μF/mm$^2$ | [40] |
| $I$ | 0.1 μA/mm$^2$ | |
| $r$ | 0.002 mm | [27] |
| $r_m$ | $10^3$ kΩ·mm$^2$ | [27] |
| $r_L$ | 1 kΩ·mm | [27] |
| $L$ | 1 mm | [27] |
| $\alpha$ | 0.65 (0.75, 0.85) | |

Figure 4 shows the well-known solution of problem (27), while figure 5 highligths the differences in the spatio-temporal distribution of the membrane potential $v(x,t)$, solution to problem (26), for $\alpha = 1$ (classic case; figure 5(a)) and for $\alpha = 0.65$ (figure 5(b)). It is apparent that near $x = 0$ the amplitude of oscillations of the membrane potential are higher for $\alpha = 0.65$ than for $\alpha = 1$, while near the node of Ranvier ($x = L$) the oscillations are higher for $\alpha = 1$. Profiles through the three-dimensional plots in figure 5 are shown in figure 6. At fixed time ($t = 0.1\ ms$), the potential increases sharper near the node of Ranvier for $\alpha = 0.65$ than for $\alpha = 1$ (figure 6(a)), while at the middle of the internodal region, the amplitude of potential's oscillations is slightly higher for $\alpha = 1$ than for $\alpha = 0.65$ (figure 6(b)).

Figures 7-10 show profiles of $v(x,t)$ in the case when the potential is assumed to be zero at the middle of the internodal revion and symmetric at the node of Ranvier. As $\alpha \rightarrow 1$, the action potential loses its sharpness at the node for the fixed time $t = 0.1\ ms$ (figure 7(a)), while its time variations are almost identical at the fixed location $x = 700\ \mu m$ (figure 7(b)). The results shown in figures 8 and 9 are obtained for $\alpha = 0.65$ (figures 8(a) and 9(a)) and for $\alpha = 1$ (figures 8(b) and 9(b)). Figure 8(a) shows that at time $t = 0.1\ ms$ the membrane potential becomes smoother at the node as the axon's diameter increases. Also, the long-range effects increase as the diamater increases: the potential is non-zero almost everywhere except in a very small neighbourhood of the middle of the internodal region. By comparison, figure 8(b) shows that in the classic case ($\alpha = 1$) the potential decays to zero away from the node regardless of the size of the axon's diameter. This suggests a local behavior. The decay becomes slower as the diameter increases. Figure 9 shows that at fixed location $x = 700\ \mu m$, the amplitude of oscillations increases with the axon's diamater in both cases ($\alpha = 0.65$ and $\alpha = 1$) which is in agreement with experimental observations. While the shapes of the oscillations are the same for the two values of $\alpha$, the amplitude is slightly higher for $\alpha = 0.65$ (figure 9(a)).

Figure 10 shows that at time $t = 0.1ms$ the shape of the membrane potential changes with the length of the internodal region for both $\alpha = 0.65$ and $\alpha = 1$, and, as expected, when $\alpha = 0.65$ the long range effects diminish as the internodal length decreases.

Figures 11-14 show the spatio-temporal distribution of the membrane potential in the bidirectional case for $\alpha = 0.65, p = 0.75, q = 0.25$ (figures 11 and 12), and $\alpha = 0.65, p = 0.25, q = 0.75$ (figures 13 and 14). The long-range effects decrease as $p$ decreases from 1 to 0.75 (figures 5, 11, and 13). The profiles of $v(x,t)$ at $t = 0.1\ ms$ are alike for $p = 1$ (figure 6(a)), $p = 0.75$ (figure 12(a)), and $p = 0.25$ (figure 14(a)). By comparing figures 6(b), 12(b), and 14(b) we see that as the value of $p$ decreases from 1 to 0.25 (and thus the value of $q$ increases from 0 to 0.75), the amplitude of oscillations of the membrane potential decreases at fixed location $x$.

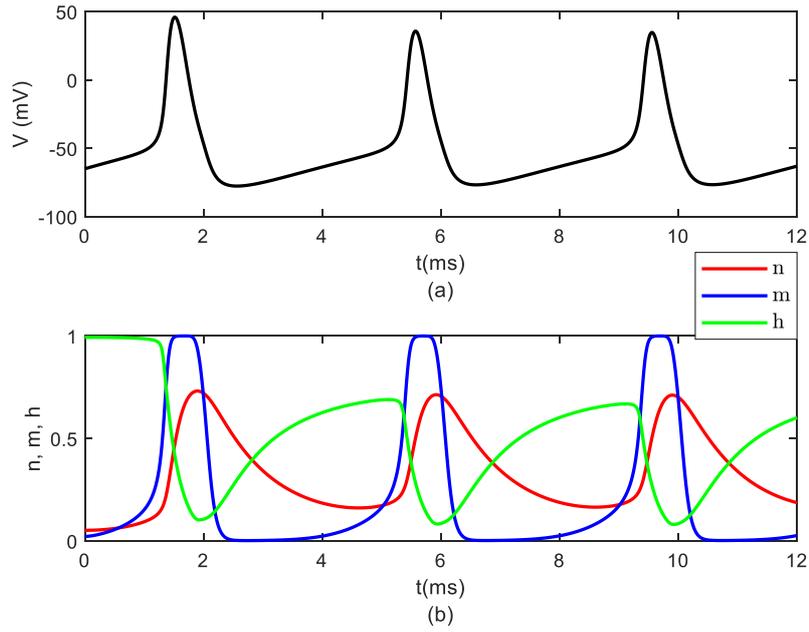

**Figure 4.** Time variations of (a) voltage $V(t)$, and (b) gating variables $n(t)$, $m(t)$, and $h(t)$, obtained by solving the modified Hodgkin-Huxley equations (27).

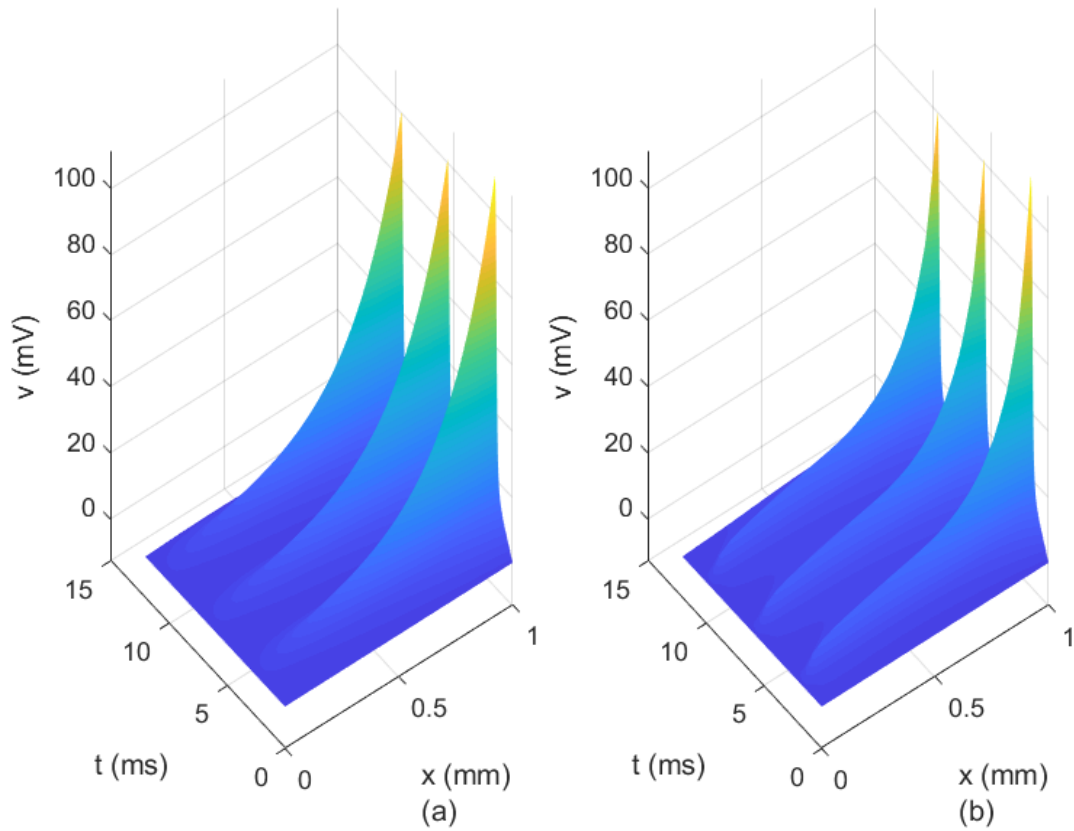

**Figure 5.** Spatio-temporal variations of voltage $v(x, t)$ for (a) $\alpha = 1$ and (b) $\alpha = 0.65$.

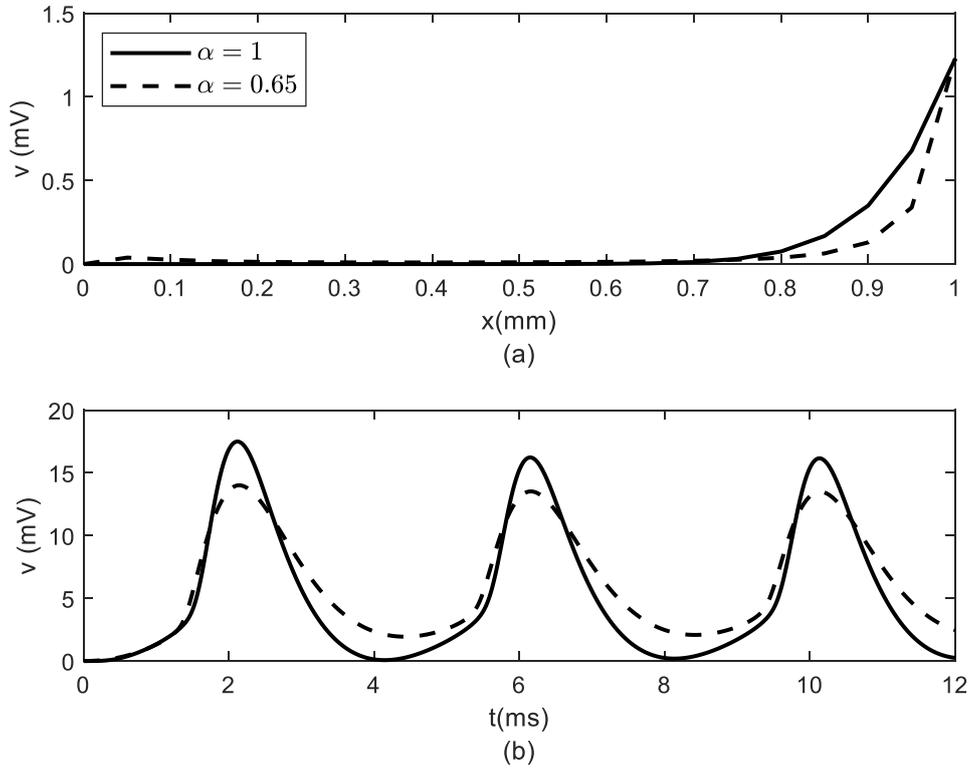

**Figure 6.** Profiles of $v(x,t)$ at (a) $t = 0.1\ ms$ and (b) $x = 500\ \mu m$ extracted from figure 5. The fixed time in part (a) is almost the time when $V(t)$ reaches its first pick (see figure 4(a)).

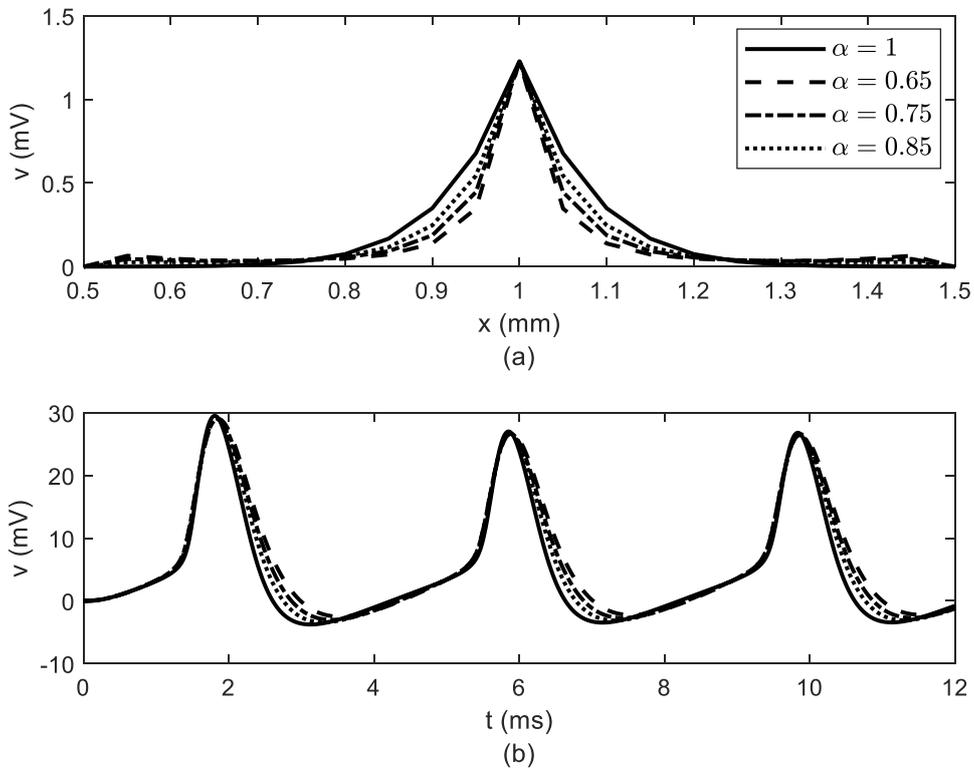

**Figure 7.** Profiles of $v(x,t)$ at (a) $t = 0.1\ ms$ and (b) $x = 700\ \mu m$ for various values of parameter $\alpha$.

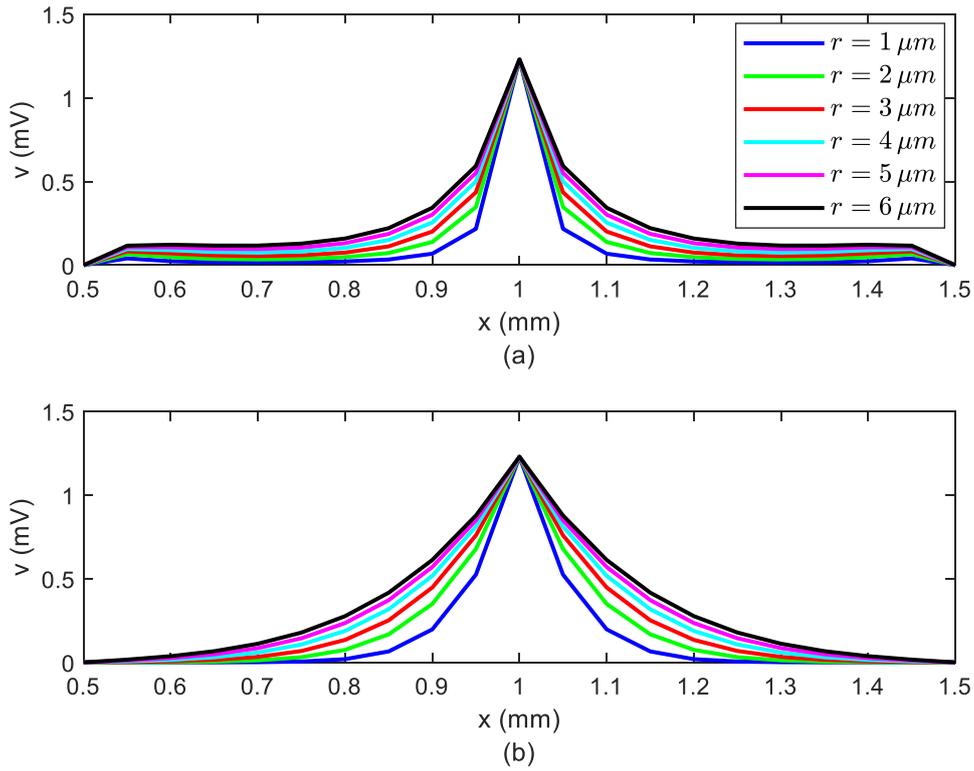

**Figure 8.** Profiles of $v(x,t)$ for (a) $\alpha = 0.65$ and (b) $\alpha = 1$ (bottom row) at $t = 0.1\ ms$ and (b) $x = 700\ \mu m$ for various values of the axon's radius $r$.

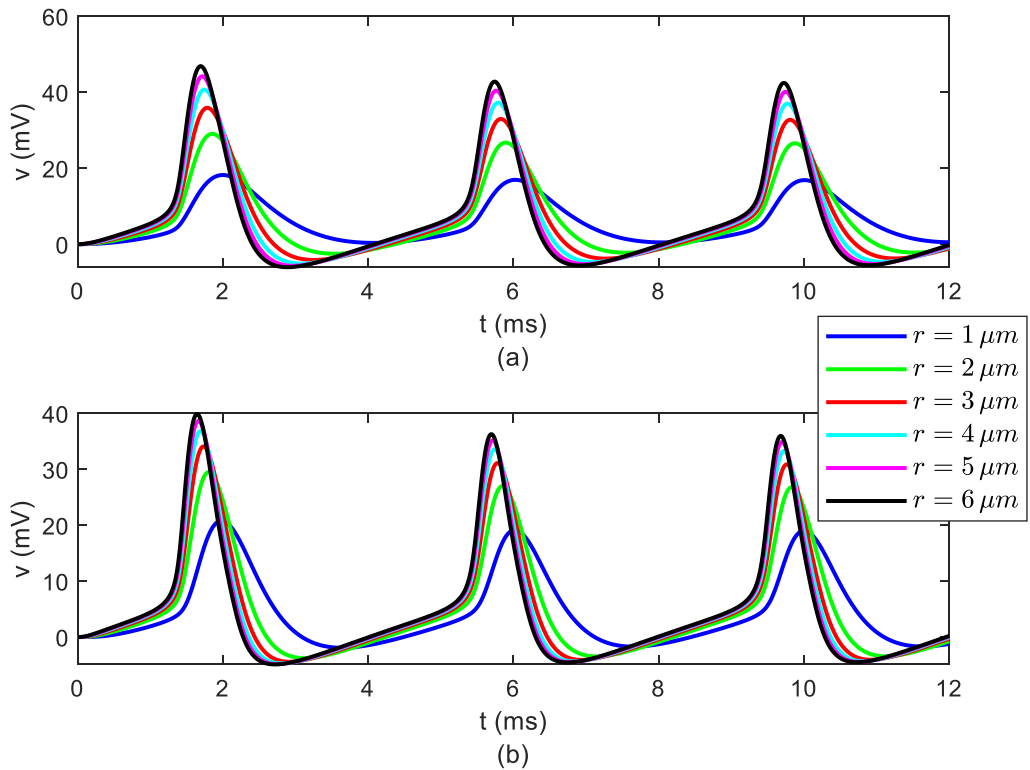

**Figure 9.** Profiles of $v(x,t)$ for (a) $\alpha = 0.65$ and (b) $\alpha = 1$ at $x = 700\ \mu m$ for various values of the axon's radius $r$.

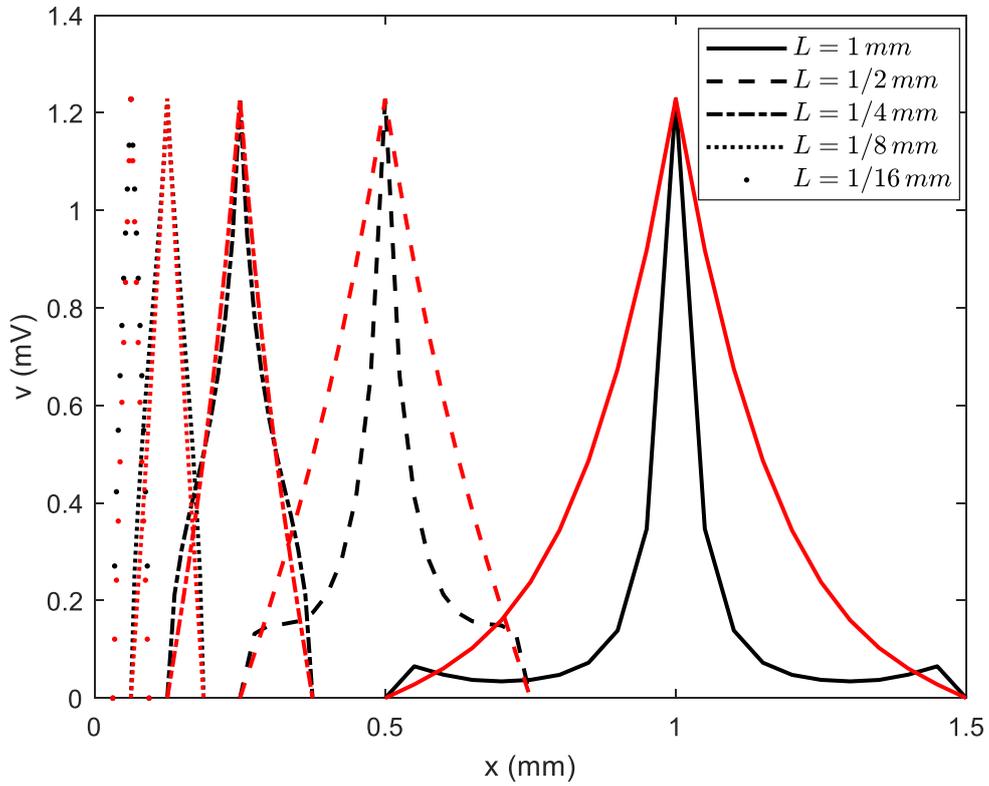

**Figure 10.** Profiles of $v(x,t)$ for $\alpha = 0.65$ (black) and $\alpha = 1$ (red) at $t = 0.1\,ms$ and for various values of the internodal length $L$. The shapes for the two values of $\alpha$ are identical for shorter internodal lengths.

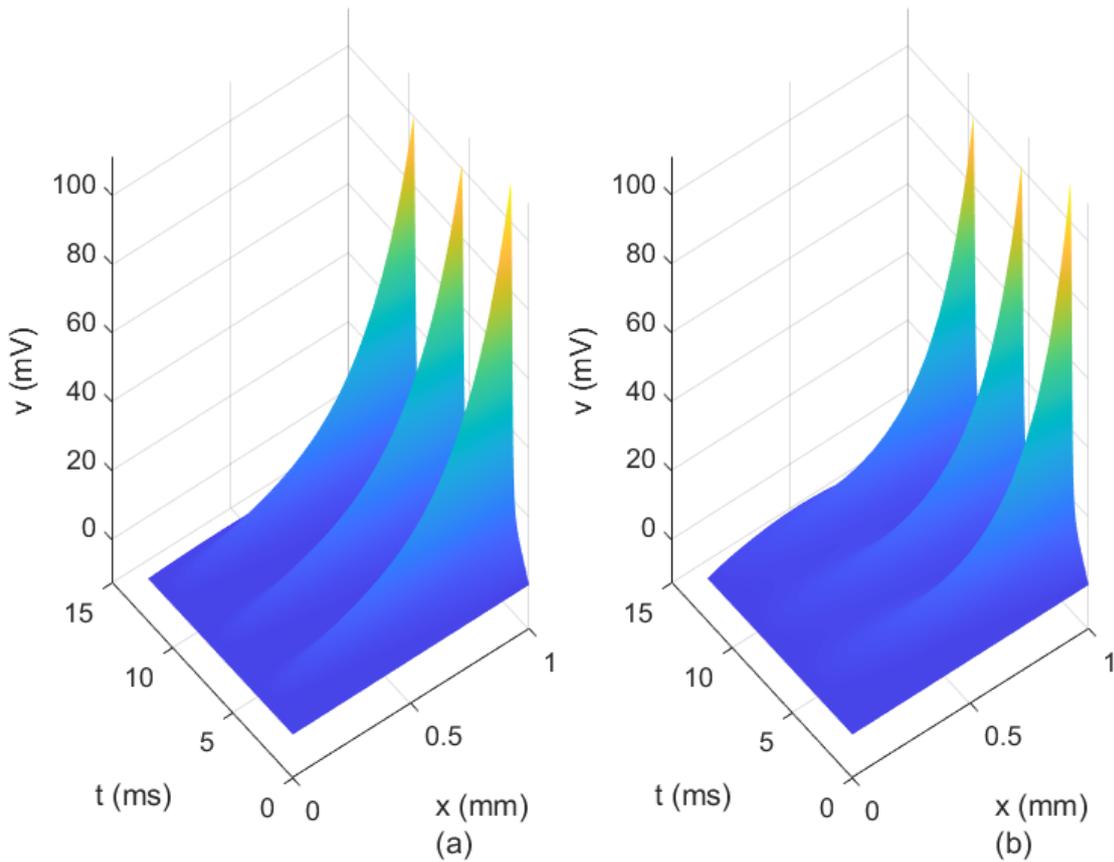

**Figure 11.** Spatio-temporal variations of voltage $v(x,t)$ for (a) $\alpha = 1, p = q = 1/2$ and (b) $\alpha = 0.65, p = 0.75, q = 0.25$.

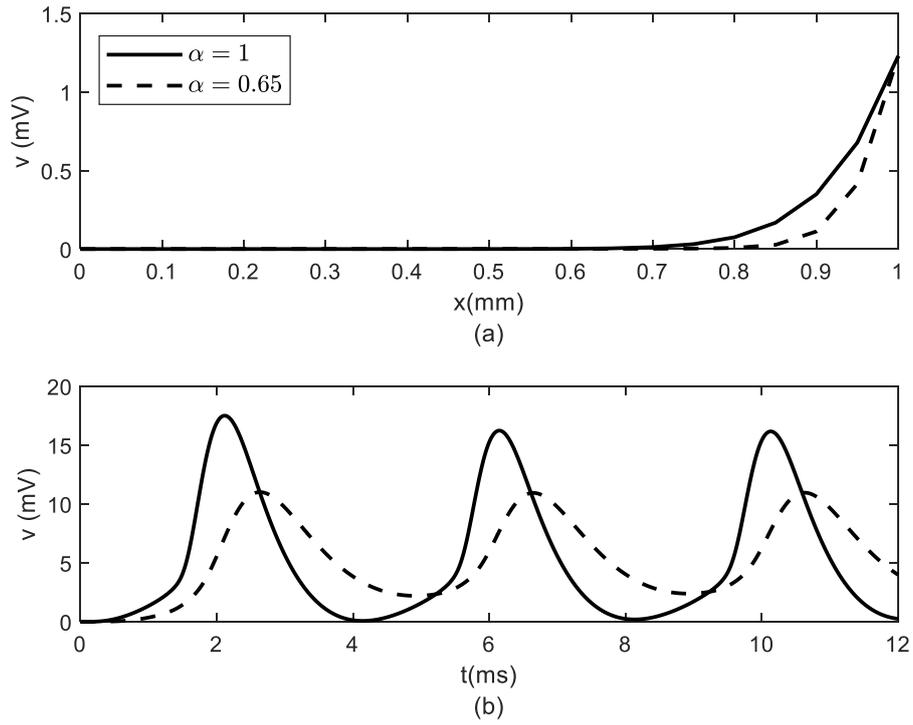

**Figure 12.** Profiles of $v(x,t)$ at (a) $t = 0.1\ ms$ and (b) $x = 500\ \mu m$ extracted from figure 11. The fixed time in part (a) is almost the time when $V(t)$ reaches its first pick (see figure 4(a)).

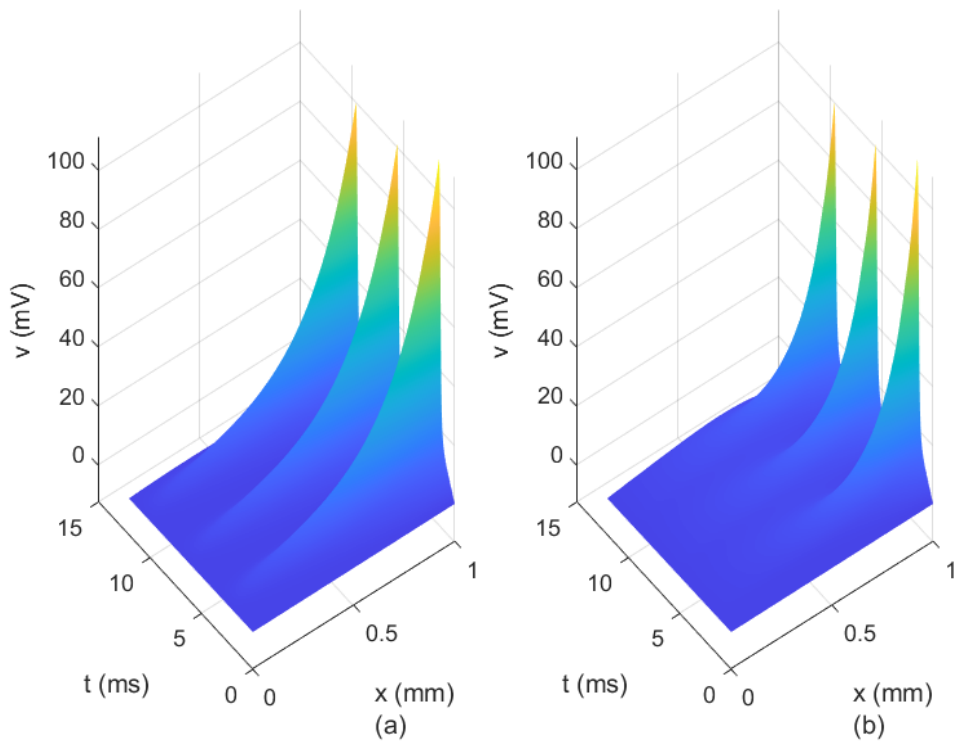

**Figure 13.** Spatio-temporal variations of voltage $v(x,t)$ for (a) $\alpha = 1, p = q = 1/2$ and (b) $\alpha = 0.65, p = 0.25, q = 0.75$.

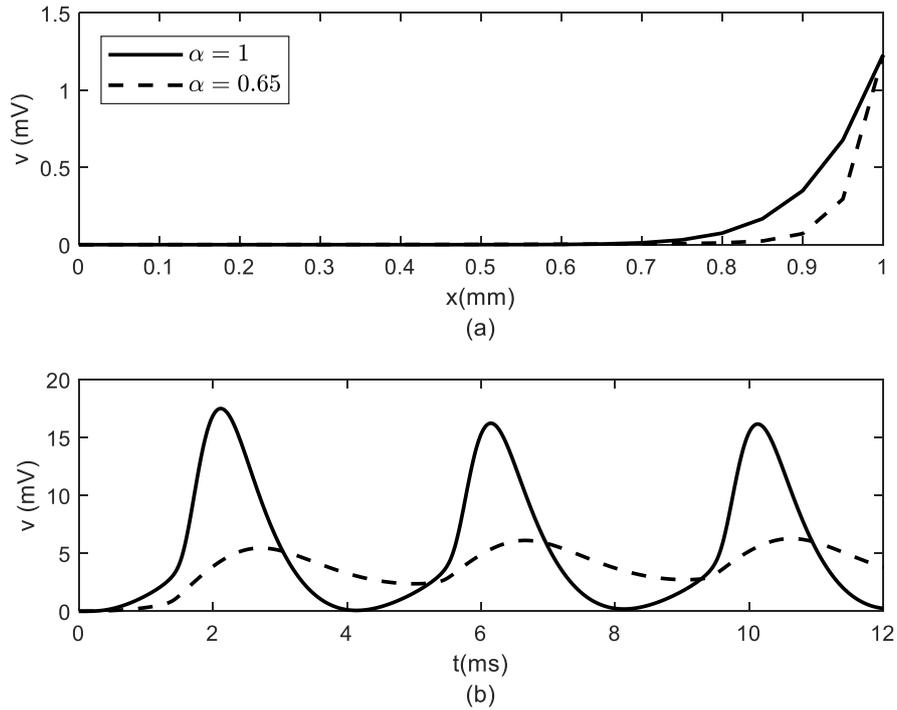

**Figure 14.** Profiles of $v(x,t)$ at (a) $t = 0.1\ ms$ and (b) $x = 500\ \mu m$ extracted from figure 13. The fixed time in part (a) is almost the time when $V(t)$ reaches its first pick (see figure 4(a)).

## 5- Conclusion

In this paper we assumed that spatial non-locality due to processes like long-range interactions of ions and water molecules at nanoscales and the anomalous diffusion of ions through the myelin sheath and the ECS affects the propagation of action potentials in myelinated neurons. We proposed a space-fractional cable equation to model spatial non-locality of the action potentials propagation. We used spatial Caputo fractional derivatives and their properties to derive this equation. Further we used a numerical method to find the spatio-temporal distribution of the membrane potential in a leaky internodal region with one isopotential node described by a modified Hodgkin-Huxley model. Numerical results were shown for this dynamic problem in the special case of unidirectional propagation of action potentials and in the case of bidirectional propagation of action potentials. In our future work we plan to explore effects on the propagation of action potentials by introducing a link between a time-dependent non-local parameter $\alpha(t)$ and the diffusion of the extracellular potassium.

## 6- References

293.